# The Collection of Didactic Physics Instruments of the Istituto Maria SS.ma del Rosario in Palermo

**Aurelio Agliolo Gallitto**

*Dipartimento di Fisica e Chimica - Emilio Segrè, University of Palermo, Palermo, Italy*

Email: aurelio.agliologallitto@unipa.it

**ABSTRACT**

In the autumn of 2025, the Istituto Maria SS.ma del Rosario (IMSSR) in Palermo donated a collection of didactic physics instruments to the Department of Physics and Chemistry - Emilio Segrè of the University of Palermo, Italy. The instruments have been identified, catalogued, classified by disciplinary area, and are now displayed within the Historical Collection of Physics Instruments, housed in the historic building at 36 Via Archirafi, Palermo. The collection, consisting of approximately sixty instruments primarily intended for teaching classical physics at the upper-secondary school programmes, represents a significant example of the experimental culture of physics education in the first half of the twentieth century. This paper reconstructs the historical origins of the collection, illustrates its disciplinary areas and the main typologies of instruments, and analyses its scientific and educational value from a descriptive and historical museological perspective, with particular attention to the role of commercial catalogues as tools of didactic and technical mediation.

**Note:** This paper is presented as a bilingual English-Italian publication. The English text is followed by the original Italian version.

.

## Introduction

The Istituto Maria SS.ma del Rosario (IMSSR) in Palermo, Italy, run by the Dominican Sisters of the Sacred Heart of Jesus, was founded on 19 June 1925 in the Dominican monastery of St Catherine in Palermo. Initially, the Institute offered a complete programme of primary and lower secondary education. Owing to the Second World War, its activities were interrupted and the school was temporarily transferred to the Palace of the Princes of Cutò in Via Lincoln, Palermo. In 1948, a plot of land was purchased at 44 Via Gian Filippo Ingrassia, where a new school building was constructed. Following the transfer to the new premises in 1949, enrolment increased and several upper-secondary curricula were established, each with two sections: Liceo Classico (Classical High School), Istituto Tecnico Femminile (Women's Technical Institute), and Perito Aziendale e Corrispondenze in Lingue Estere (PACLE, Business Administration and Foreign Languages Correspondence). The teaching laboratories were subsequently expanded, and numerous scientific instruments were acquired. During the same period, the convent school accommodated more than 150 girls from neighbouring villages. Over the years, however, the management of the Institute became increasingly complex, eventually leading to the closure of upper-secondary programmes. In the autumn of 2025, following the discontinuation of laboratory activities, the Institute donated its physics and chemistry teaching instruments to the Department of Physics and Chemistry - Emilio Segrè (DiFC) of the University of Palermo.

The instruments were identified, classified into disciplinary sections, and displayed in two period display cases within the Historical Collection of Physics Instruments (website 1), housed in the historic building at 36 Via Archirafi. Figure 1 shows the two display cases. The collection, consisting of approximately sixty instruments primarily intended for the teaching of classical physics, has been named the IMSSR Collection. A complete list of the instruments, accompanied by brief descriptions, is available on the dedicated web page (website 2).

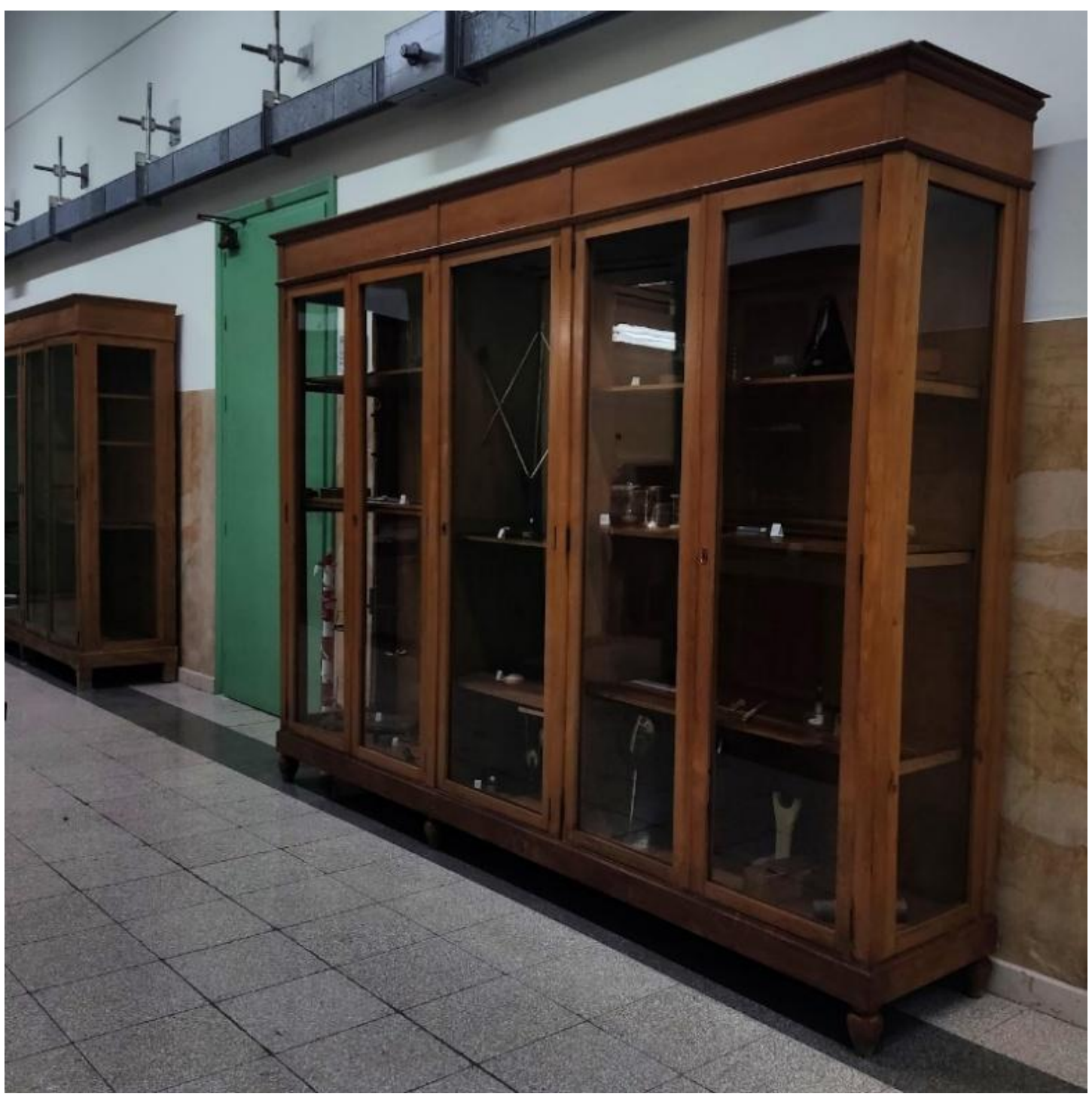

**Figure 1.** Display cases installed in the historic building at 36 Via Archirafi, DiFC, University of Palermo.

The IMSSR Collection constitutes a significant testimony to the experimental culture of school physics in the first half of the twentieth century, documenting a historical period in which physics teaching in Italian secondary schools relied largely on the systematic use of standardized experimental apparatus selected from commercial catalogues (Frick, 1856; Pike, 1856; Fumeo, ca. 1930; Paravia, 1938 and 1953; Magini, 1940; Vallardi, 1960). These instruments reflect a strongly experimental approach to physics education, consistent with the tradition of classical physics outlined in nineteenth-century textbooks and maintained in school curricula until the mid-twentieth century (Ganot, 1863; Jamin, 1870; Murani, 1917).

This article illustrates the disciplinary areas and the principal categories of instruments represented in the IMSSR Collection. It analyses several instruments of particular interest as documentary sources for the reconstruction of the history of physics education, a field that has received less scholarly attention than scientific instrumentation intended for research. From a museological perspective, such collections make it possible to examine the relationships among scientific knowledge, education, and industrial production, highlighting the role of commercial catalogues as instruments of educational and technical mediation.

## Disciplinary Areas of the Collection

The instruments were classified according to disciplinary criteria consistent with the teaching tradition of classical physics. The most extensively represented area is mechanics, with instruments intended for the study of motion, equilibrium, and simple machines, including inclined planes, pulleys, and dynamometers. Figure 2 shows an apparatus for demonstrating the parallelogram of forces, equipped with two pulleys and four graduated rods. The apparatus is described by Frick in his 1856 catalogue (Frick, 1856: 58) and illustrated in Figure 3.

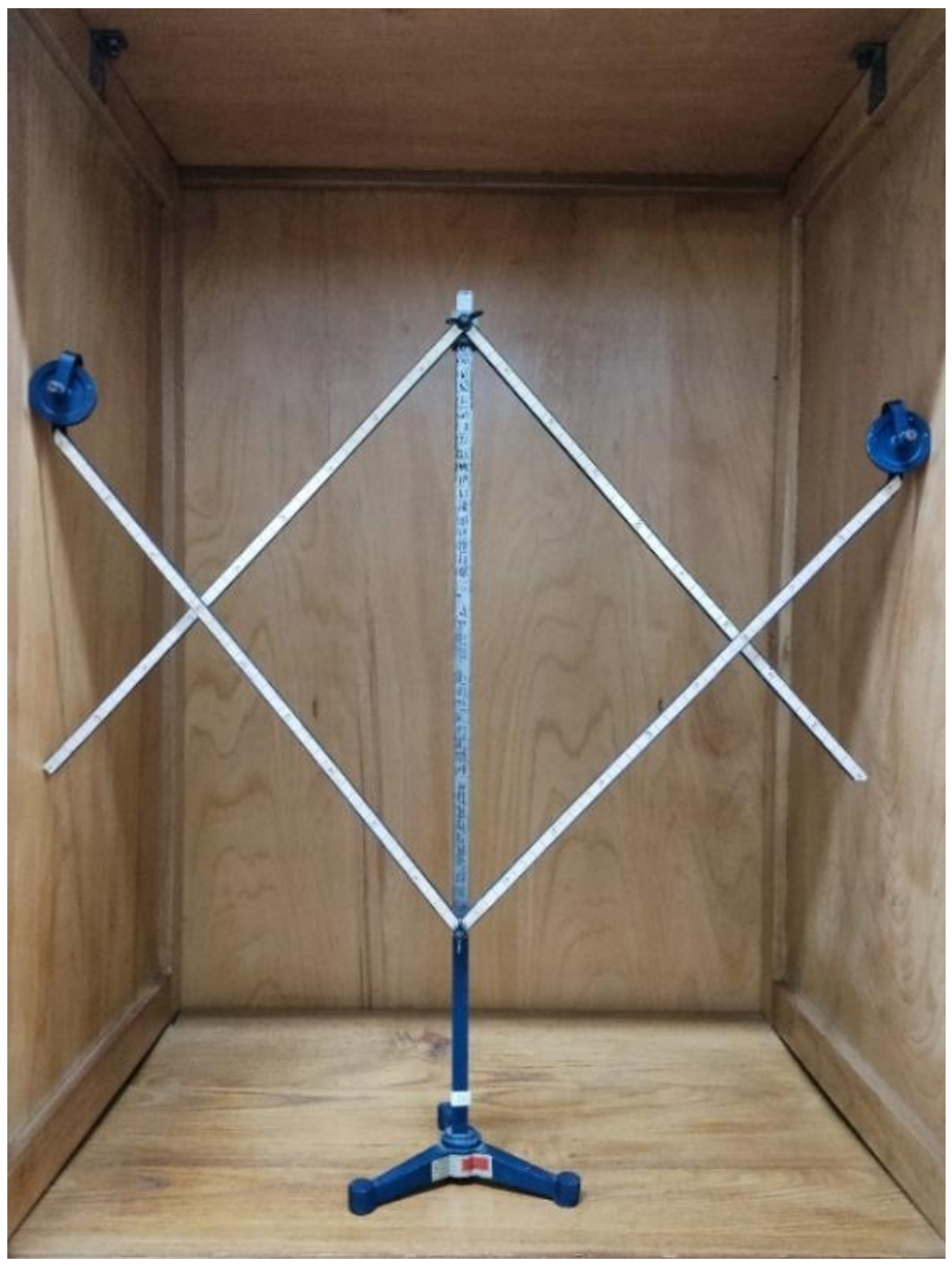

**Figure 2.** Apparatus for the parallelogram of forces, with two pulleys and four graduated rods, manufactured by Vallardi, ca. 1950.

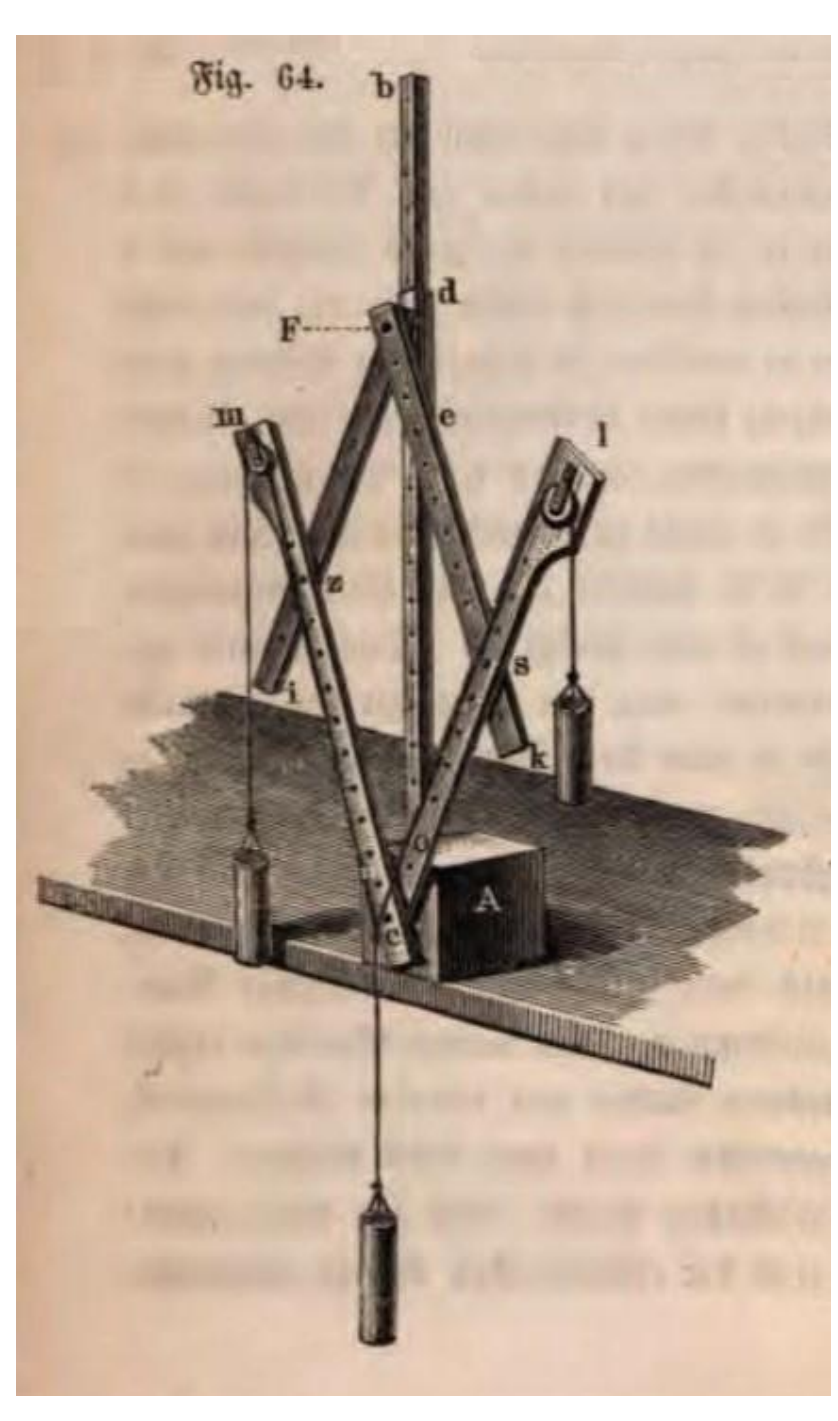

**Figure 3.** Diagram of the apparatus for the parallelogram of forces, with two pulleys and four graduated rods, reproduced from Frick (1856: 58).

The section devoted to the study of fluids includes instruments designed to demonstrate the principal laws of hydrostatics and pneumatics, widely employed in physics teaching in secondary schools between the late nineteenth century and the first half of the twentieth century. Among the most common types are instruments for studying pressure in liquids, hydrostatic thrust, and Archimedes' principle, as well as devices demonstrating the transmission of pressure in fluids and phenomena related to atmospheric pressure.

Figure 4 shows a hydrostatic balance with a double Archimedes cylinder; the same figure also includes a constant-volume Nicholson hydrometer, used to determine the specific gravity of solid bodies insoluble in water. The use of this instrument is described in the extract from Pike's catalogue (Pike, 1856: 227) reproduced in Figure 5. Such devices made it possible to visualise abstract concepts through simple and robust arrangements specifically designed for repeated educational use.

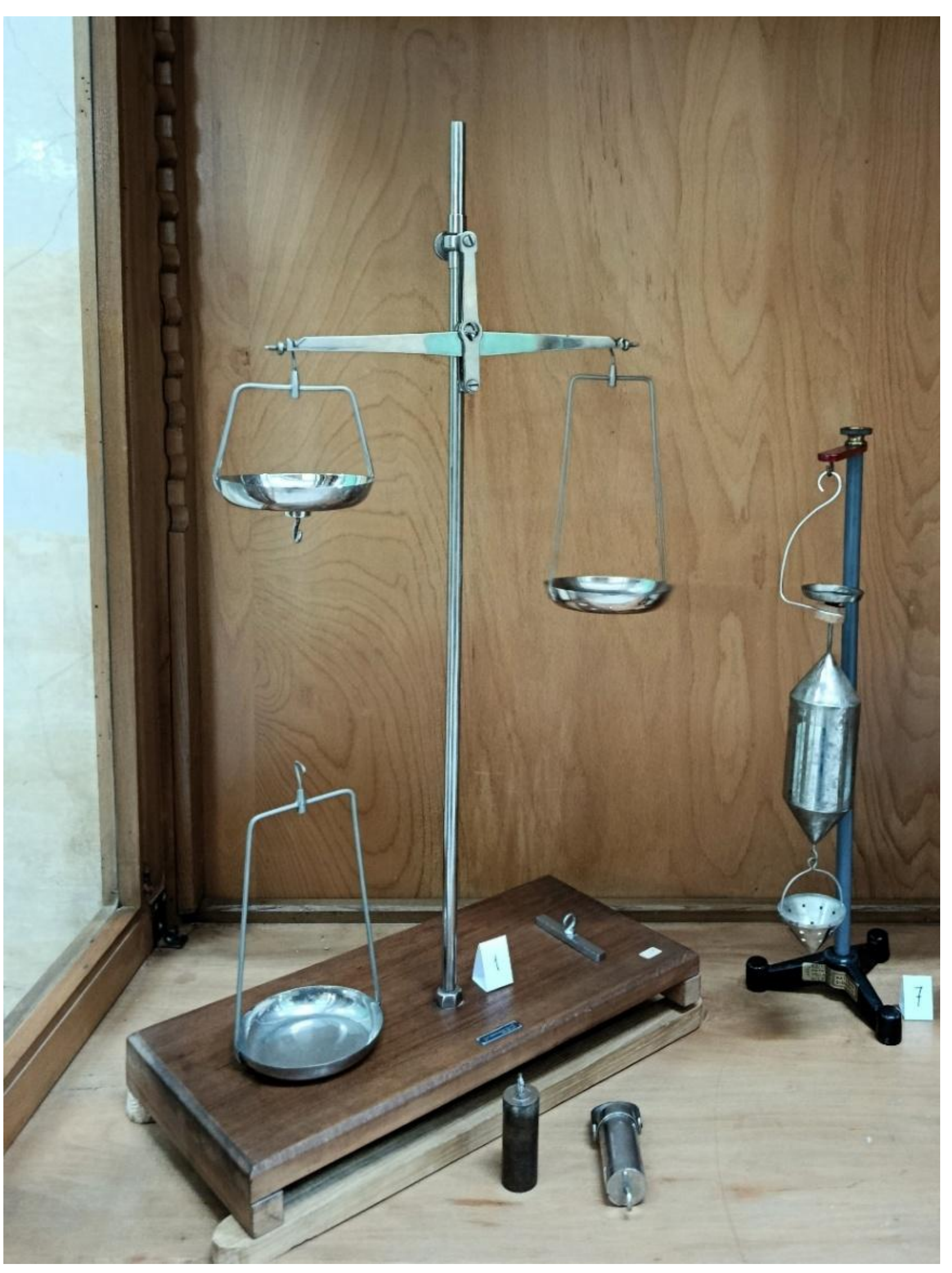

**Figure 4.** Elementary hydrostatic balance with double Archimedes cylinder mounted on a wooden base, with an adjustable beam regulated by a pressure screw; it includes three pans of equal weight, one of which is fitted with a hook for suspending the double Archimedes cylinder (G. B. Paravia Catalogue, 1953, No. 50120, Lire 9,800). On the right is a constant-volume Nicholson hydrometer for determining the specific gravity of solid bodies insoluble in water.

HYDRAULICS AND HYDROSTATICS. 227

*Nicholson's Portable Balance.*—(Fig. 241.)—This consists of a hollow cylinder of metal terminated at each end by a cone; from the top of the upper cone rises a small stem of brass, terminated by a cup, *a*, which may be slipped off; from the end of the lower cone is suspended a cup, *e*, that is loaded with a weight, sufficient to sink the instrument in water nearly to the base of the upper cone; by placing weights in the cup, *a*, at the top, it is immersed to the mark on the stem; if the weights be removed and a solid body, whose weight is less than theirs, placed in the cup, it may be immersed again to the same mark, by placing weights along with the solid in the cup, and thus the weight of the solid obtained; now place the solid in the cup attached to the weight and weigh it in water; having now the weight of the body in air, and its loss of weight in water, its specific gravity may be ascertained by dividing the weight in air by the loss of weight in water, the quotient is the specific gravity. Price, $3.50.

Fig. 241. Fig. 242.

a

e

*Nicholson's Hydrostatic Balance.*—(Fig. 242, as above.)—This instrument is made of brass, highly finished, and having a cage over the counterpoise for confining any solid body that is lighter than water. Price, $6.00.

**Figure 5.** Description and illustration of Nicholson's hydrometer from Pike (1856).

The thermology section includes instruments intended for the study of thermal phenomena, expansion, and heat transfer. Figure 6 shows an Ingenhousz apparatus for comparing the thermal conductivity of five different substances (aluminium, brass, plexiglass, steel, and wood). Also noteworthy is the presence of a rare Six-Bellani maximum-minimum thermometer (Jamin, 1870: 105), originally invented in 1782 by James Six (1731-1793) and improved in the early nineteenth century by Angelo Bellani (1776-1852).

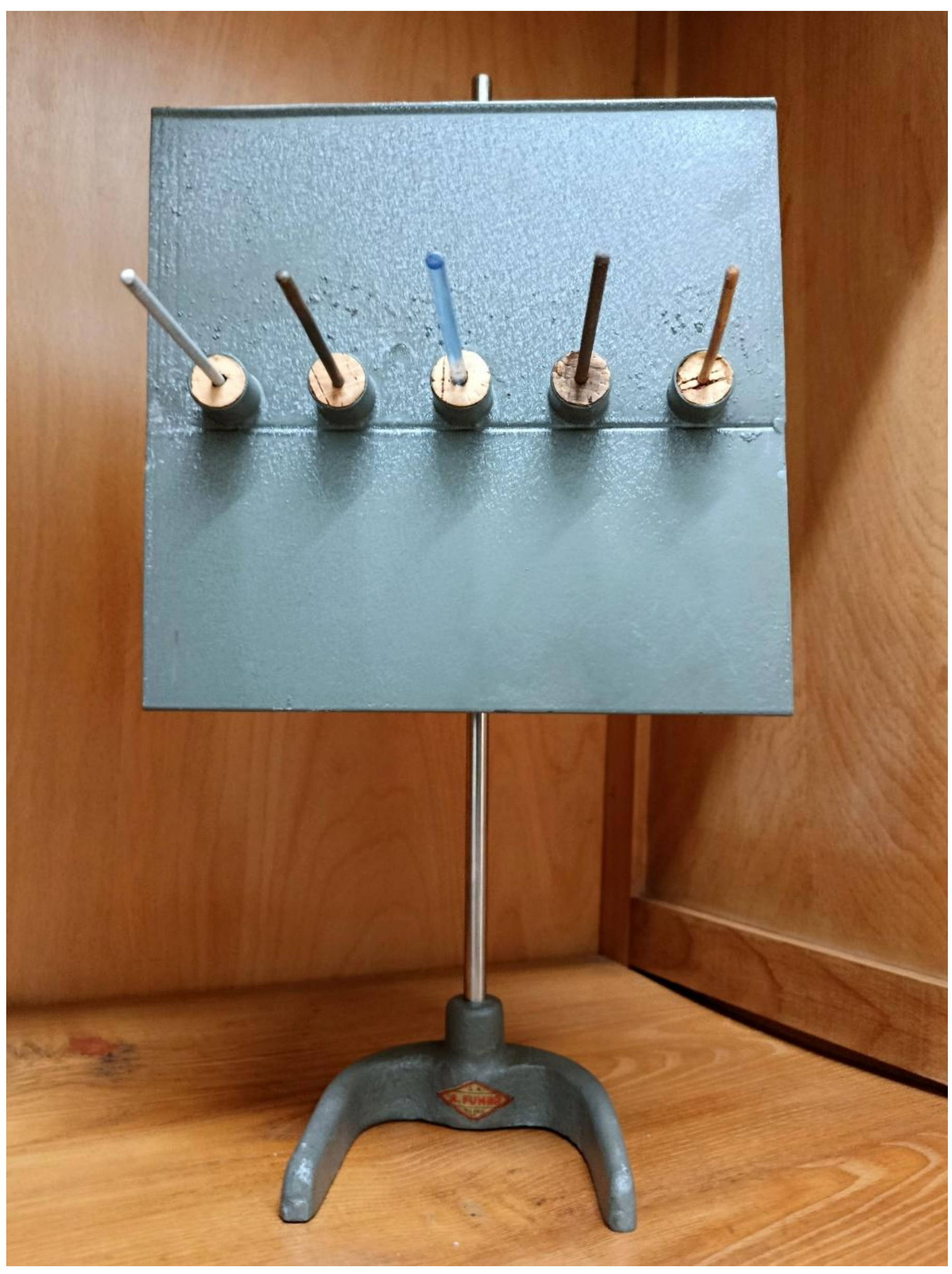

**Figure 6.** Ingenhousz apparatus for comparing the thermal conductivity of five substances (aluminium, brass, plexiglass, steel, and wood), represented by rods fitted into a sheet-metal box, manufactured by Fumeo, ca. 1950.

The acoustics instruments, including a Maelzel-Paquet metronome, a tuning fork mounted on a resonance box, and an apparatus for studying the propagation of sound in a vacuum, testify to a pedagogical approach centred on sensory experience and the relationship between physical phenomena and auditory perception. Such instruments are widely described in physics manuals and catalogues intended for educational purposes during the nineteenth and early twentieth centuries (Blaserna, 1875).

The optics section includes instruments employed in the study of the fundamental phenomena of light propagation, reflection, and image formation. The preserved apparatus reflects a didactic approach strongly oriented toward geometrical optics. Particular importance is attached to instruments designed to demonstrate the laws of reflection, both in simple configurations and in more complex arrangements intended to visualise phenomena such as multiple reflection and the dependence of images on the orientation of reflecting surfaces. The use of plane, concave, and convex mirrors allowed intuitive exploration of concepts such as focal points and the formation of real and virtual images. More popular and demonstrative instruments, such as the kaleidoscope, also helped stimulate students' interest through the observation of readily visible optical and geometrical effects.

The section devoted to magnetism and electrology includes instruments used to study magnetic and electrical phenomena through predominantly qualitative demonstrations. The electrical section documents the teaching of electrostatic charging, electrical storage and discharge, and the first notions of electric circuits. Instruments such as electroscopes, electrophorus devices, and Leyden jars were widely employed to demonstrate the presence of electric charges and the mechanisms of induction, providing visually striking yet highly effective educational experiences. Alongside these, simple circuit components and electrical generators testify to the gradual introduction of applied electricity into school teaching.

With regard to magnetism, the collection includes instruments intended for detecting and studying magnetic fields, the orientation of magnetic needles, and the interactions between magnets and electric currents. Compasses, magnetic needles, and demonstration devices made it possible to illustrate intuitively concepts such as magnetic fields, polarity, and the magnetic action of electric currents. More sensitive instruments, such as galvanometers equipped with an astatic system, enabled the detection of weak currents and introduced students to the quantitative study of electromagnetic phenomena.

For this purpose, the astatic system of the apparatus shown in Figure 7 reproduces the so-called *multiplier* of Nobili's astatic galvanometer, devised in 1825 by the physicist Leopoldo Nobili (1784-1835), from whom the instrument takes its name (website 3). To eliminate the influence of the Earth's magnetic field, Nobili used two identical needles rigidly mounted parallel to one another on the same axis but with opposite magnetic orientations and suspended from a single silk fibre. This double-needle arrangement significantly increased the sensitivity of the galvanometer compared with single-needle instruments.

Figure 8 illustrates the astatic system as reproduced in Ganot (1863):

> "*... it consists of two needles placed parallel one above the other, with opposite poles facing each other (Fig. 653), one inside and the other outside the circuit, connected in such a way that neither can rotate independently of the other [...] the principal effect of the astatic system is to reduce the directing action of the Earth.*"

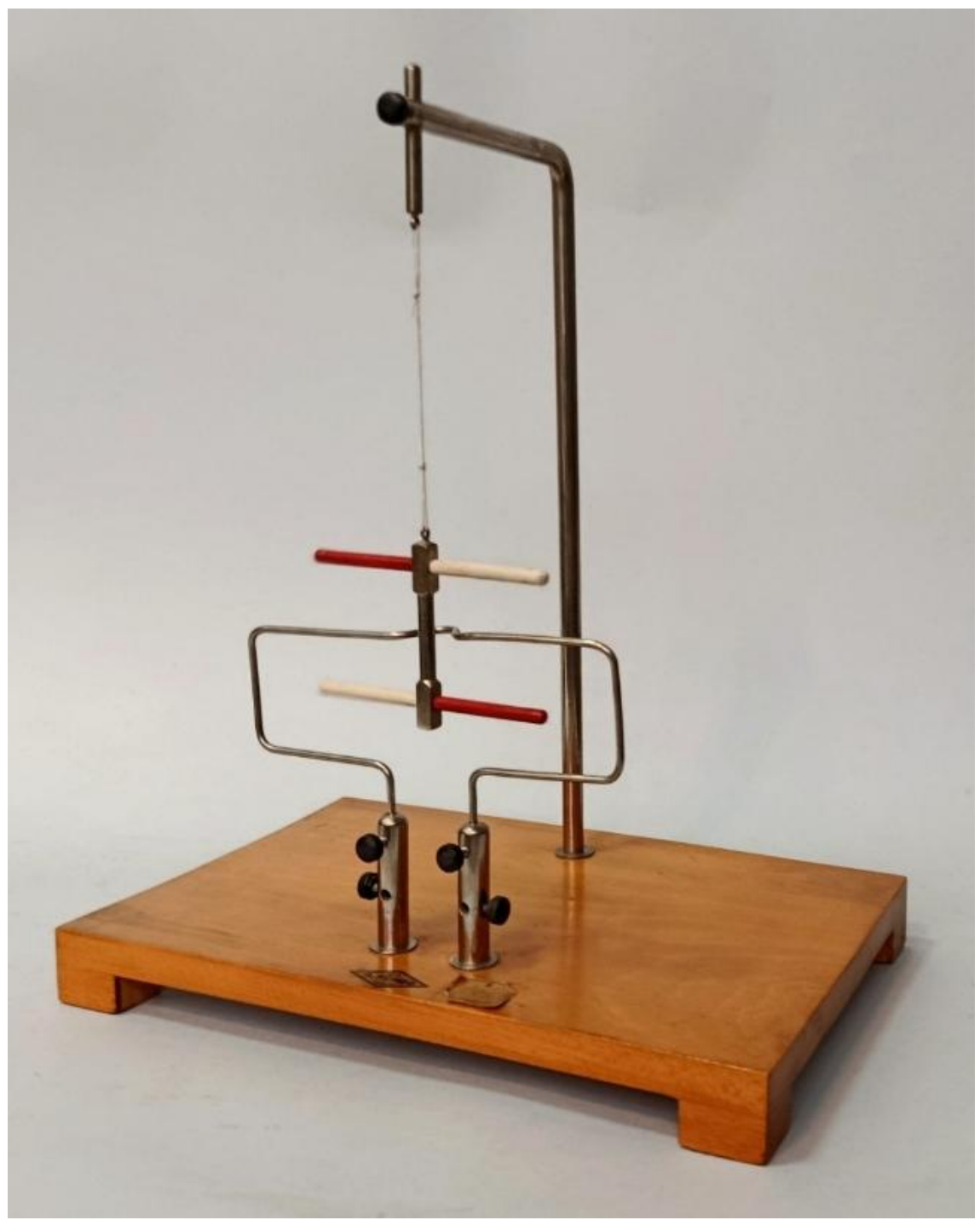

**Figure 7.** Astatic system of Nobili's galvanometer, manufactured by Fumeo, ca. 1950.

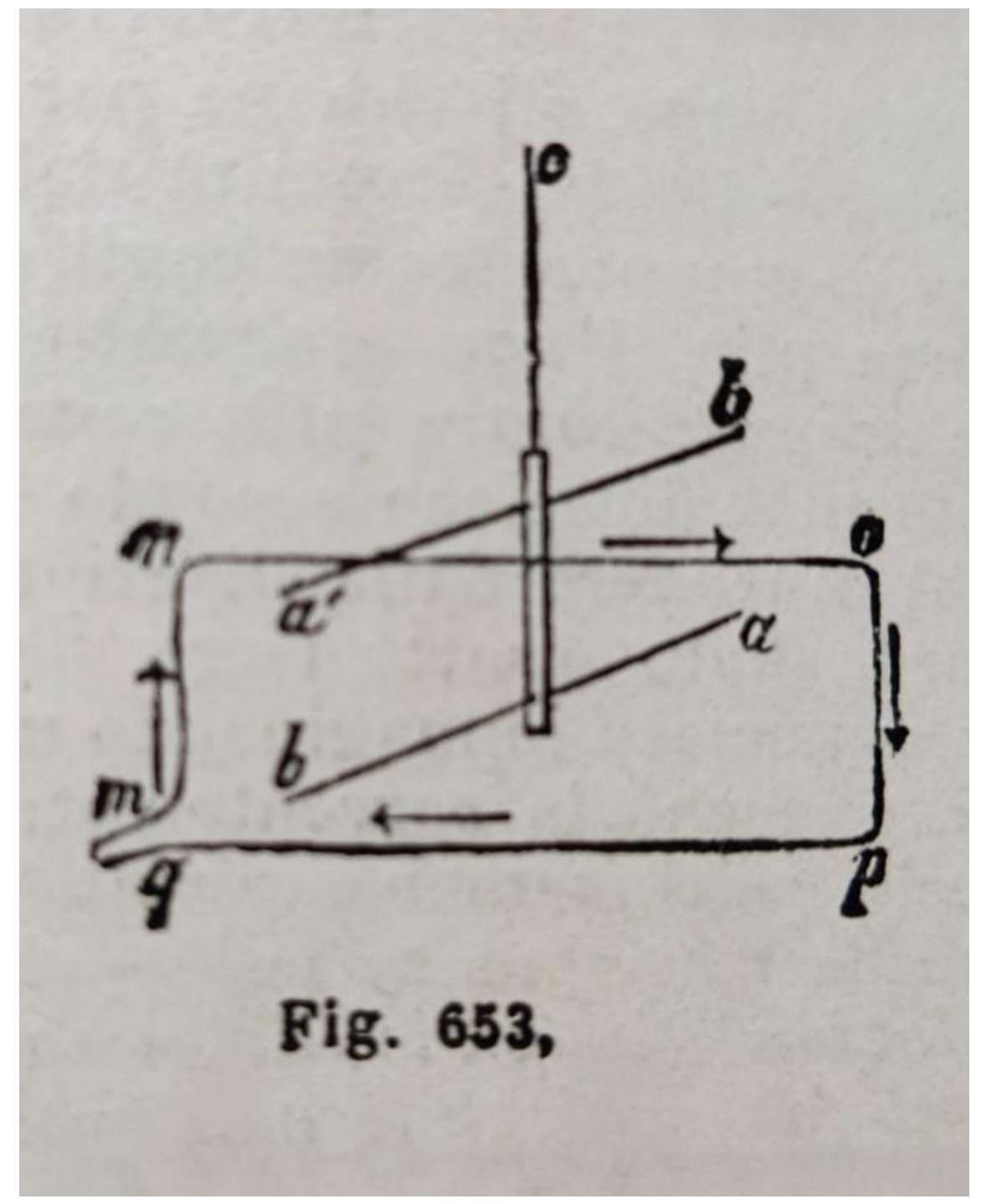

**Figure 8.** Diagram of the astatic system with multiplier of Nobili's galvanometer (Ganot, 1863: 602).

Finally, the chemistry and electrochemistry section includes instruments intended for the experimental teaching of the fundamentals of general chemistry and electrochemical phenomena, used in school laboratories during the first half of the twentieth century. The preserved apparatus illustrates a pedagogical approach strongly based on direct manipulation, observation of phenomena, and the controlled reproduction of chemical and electrolytic reactions.

Figure 9 shows a Kipp apparatus, widely used in laboratories for the production of hydrogen, hydrogen sulphide, and carbon dioxide through suitable reactions between an aqueous acid solution and an insoluble solid substance (either a metal or a salt). The apparatus was first developed in 1844 by the Dutch chemist Petrus Jacobus Kipp (1808-1864). It generally allowed the production of various gases, in controllable quantities, through the reaction between an aqueous acid solution and an insoluble solid.

Particularly noteworthy are the instruments designed for learning chemical nomenclature and the composition of substances, reflecting the need to make abstract concepts accessible through material supports and combinatorial models. An example is the FORMULATOR chemical nomenclature kit, containing 72 wooden discs, each bearing on one side the symbol of a chemical element, colour-coded according to valence. The discs could be connected by means of small elastic metal rods, allowing the construction and visualisation of even complex chemical formulae (Paravia Catalogue 1953, No. 50265, Lire 6,800). Devices of this kind highlight the importance attributed in school education of the period to visualising relationships among chemical elements and to constructing chemical formulae as both a logical and manual exercise.

The section also includes apparatus intended for the study of electrochemical processes, such as the generation of electric current through chemical reactions and the electrolytic decomposition of solutions. Instruments such as batteries and voltameters allowed students to explore the relationship between chemistry and electricity, providing concrete examples of energy transformation and introducing fundamental principles of electrochemistry.

Alongside these instruments, the presence of laboratory glassware testifies to the emphasis placed on teaching basic experimental techniques. In particular, gas-producing devices such as the Kipp apparatus (Erdmann, 1900; website 4), shown in Figure 9, made it possible to conduct fundamental demonstrations concerning the properties of gases, an essential component of scientific education at the time.

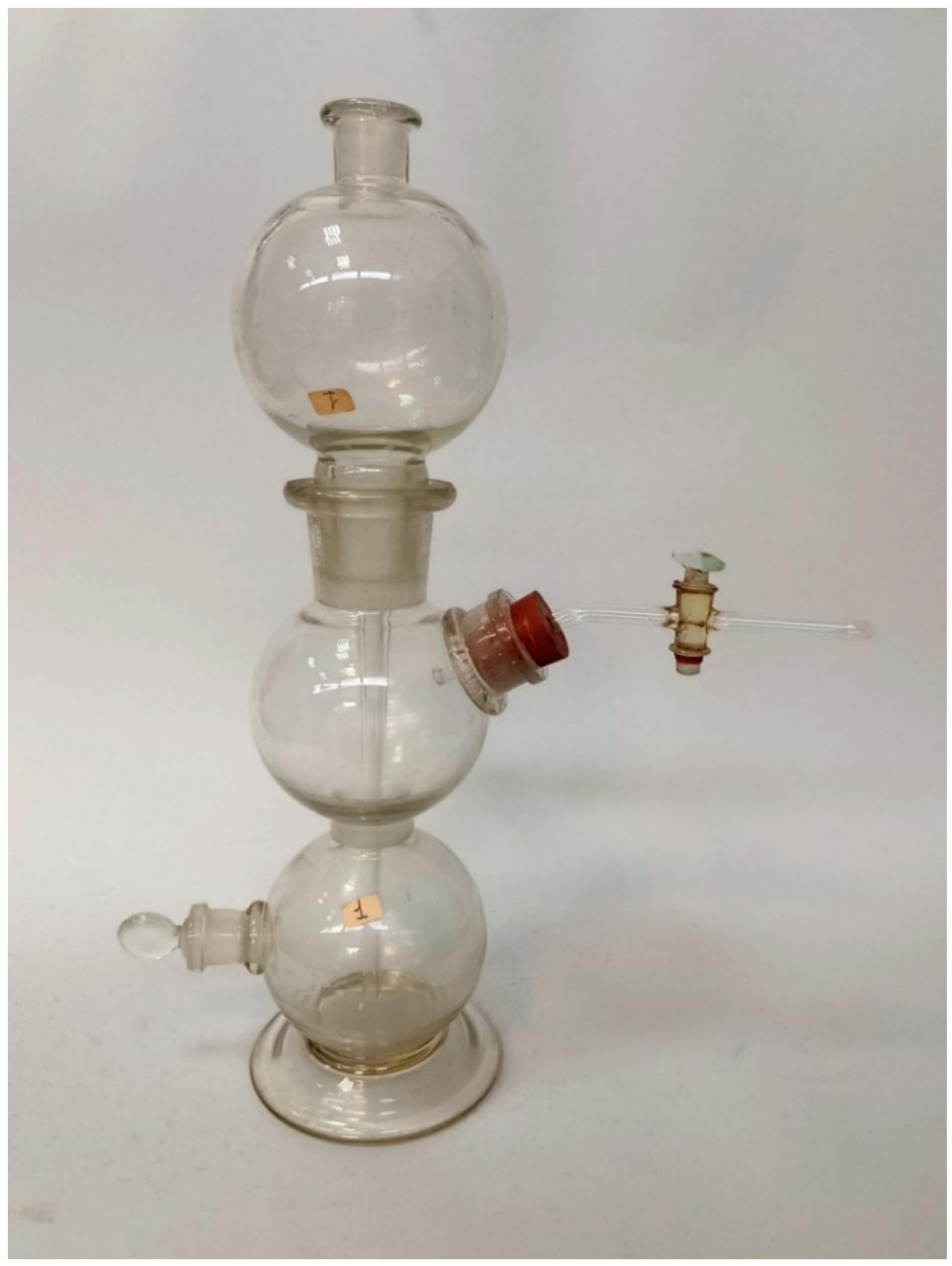

**Figure 9.** Kipp apparatus. Developed in 1844 by the Dutch chemist Petrus Jacobus Kipp (1808-1864) for the production of gases (such as hydrogen, carbon dioxide, and hydrogen sulphide) through the reaction of an acid with a metal or a salt.

A particularly significant aspect of the IMSSR Collection is the close correspondence between the preserved instruments and the commercial catalogues of scientific educational equipment manufacturers. Catalogues produced by firms such as Amedeo Fumeo, Giovan Battista Paravia, and Antonio Vallardi were not merely sales tools but also played a central role as cultural mediators, helping to guide educational choices and standardise the equipment of school laboratories.

Figure 10 shows the manufacturers' labels found on the instruments. Through the selection, description, and classification of apparatus, these catalogues contributed to defining educational practices by suggesting methods of use, experimental sequences, and disciplinary contexts of application. Consequently, they constitute a privileged documentary source for studying the relationship between industrial production, school curricula, and experimental teaching, as well as for analysing the processes through which scientific knowledge circulated and became established within schools.

From this perspective, the IMSSR Collection makes it possible to establish a direct relationship between scientific instruments and the published sources that promoted their dissemination, offering an integrated interpretation of the material culture of science education in Italian schools during the first half of the twentieth century.

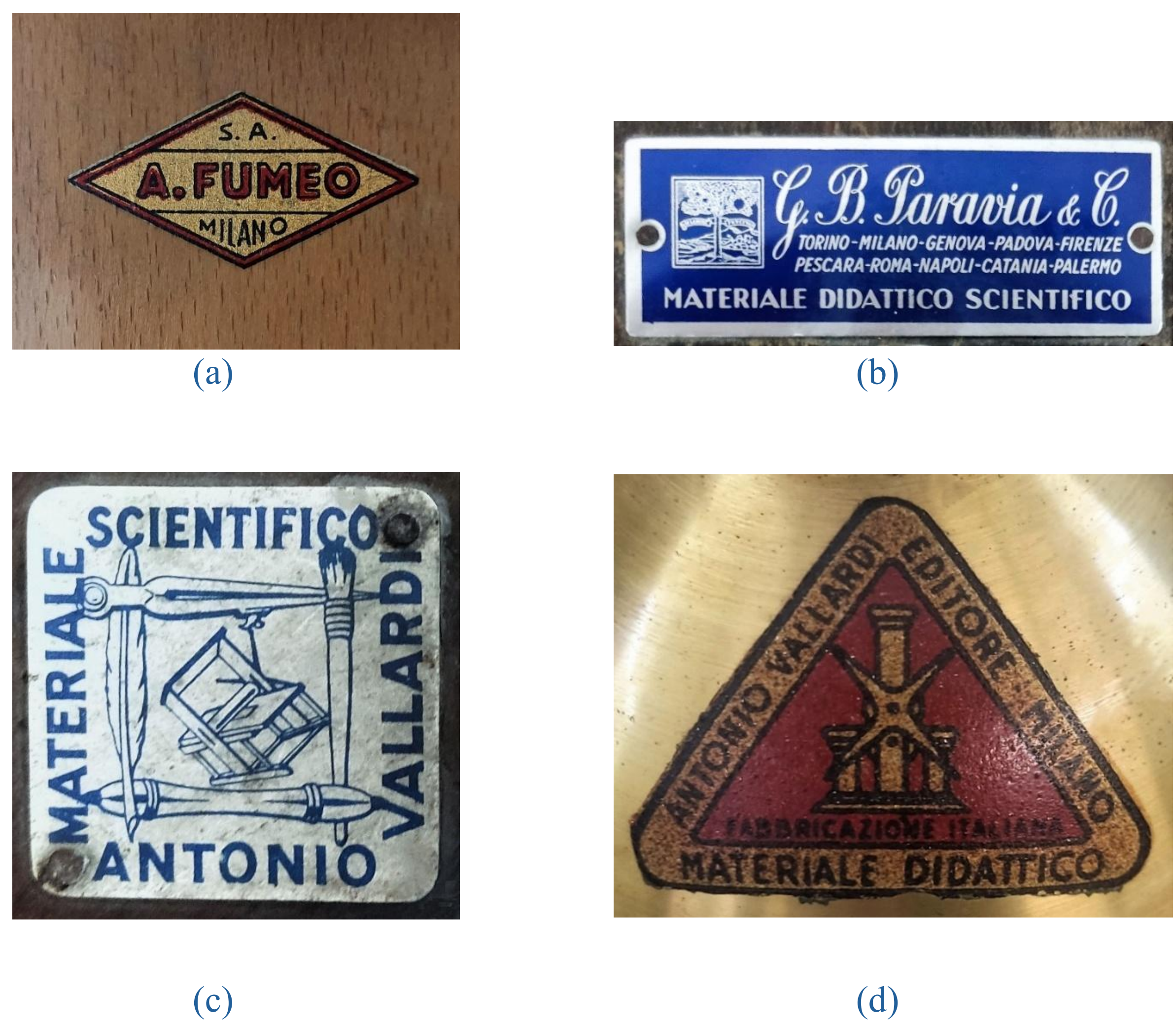

**Figure 10.** Manufacturers' labels found on instruments of the IMSSR Collection: (a) Amedeo Fumeo, Joint-Stock Company, Milan; (b) Giovan Battista Paravia & Co., Scientific Educational Equipment, Turin; (c) Antonio Vallardi, Scientific Equipment, Milan; (d) Antonio Vallardi, Publisher, Milan.

## Discussion and Conclusions

The instruments of the IMSSR Collection reflect a pedagogical approach based on the direct observation of physical phenomena and the experimental demonstration of fundamental laws, in accordance with a tradition firmly established in school practice between the late nineteenth century and the first half of the twentieth century. The analysis of the different disciplinary sections reveals common educational models based on the use of simple, robust, and standardized apparatus designed for repeated use in school laboratories.

The case of the IMSSR Collection contributes to filling a gap in the literature on school scientific instrumentation, which has often been neglected in comparison with academic and research instruments. From a historical and educational perspective, the instruments document a period in which physics teaching in Italian secondary schools relied extensively on the systematic adoption of laboratory instruments manufactured in series and distributed through the commercial catalogues of the principal firms specializing in scientific educational equipment. The close correspondence between the preserved instruments and their catalogue descriptions testifies to the central role played by such catalogues in shaping educational practices and standardizing laboratory equipment. From this standpoint, the IMSSR Collection constitutes a privileged documentary source for reconstructing the experimental culture of school physics, highlighting the relationship among scientific knowledge, experimental teaching, and industrial production. The transfer of these instruments from their original educational context to a museum setting now allows them to be interpreted from a historical and comparative perspective, enhancing their value not only as teaching aids but also as documents of the history of science education.

The transfer of the collection from the school environment to the museum context has therefore changed its status, transforming the instruments from functional objects into cultural heritage assets of scientific interest. Their integration into the Historical Collection of Physics Instruments of the University of Palermo (Sear, 2017; Agliolo Gallitto et al., 2018) makes it possible to situate them within a broader historical and scientific framework, in relation both to contemporary university instruments and to the educational and research practices developed in academic settings (Barbacci et al., 2012; Agliolo Gallitto et al., 2017). In this context, the IMSSR Collection constitutes a coherent and well-integrated group of instruments within the Historical Collection of Physics Instruments of the University, capable of providing a comprehensive view of the teaching of physics and related sciences in Italian schools during the first half of the twentieth century, while highlighting both continuities and differences between school and university education. The donation has substantially enriched the university collection, which today comprises approximately six hundred scientific and educational instruments dating from the early nineteenth century onwards (Sear, 2017; Agliolo Gallitto et al., 2018; website 1).

From a museological perspective, the enhancement of the IMSSR Collection fully accords with the objectives of scientific museology, particularly with regard to the preservation, interpretation, and transmission of the material culture of science. Its incorporation into the university heritage has made possible not only the conservation of the instruments but also their historical contextualization, systematic cataloguing, and accessibility for research, teaching, and public outreach purposes.

In conclusion, the IMSSR Collection represents a significant example of a historical collection of scientific instruments of school origin. Its study contributes to integrating the history of physics, the history of science education, and descriptive and historical museology, offering new perspectives for research on scientific heritage and for reflection on the practices of preservation and enhancement of educational collections within university contexts.

## Acknowledgements

The author wishes to thank Sister Rosa Maria La Bella for making the donation possible; Maria Rosalia Carotenuto, Salvatore Cataldo, and Fabio Panfili for their valuable suggestions; Marcello Mirabello for technical support; and Germana Mulè and Francesca Tignola for their assistance in researching of historical bibliographical sources.

## Websites (last accessed 22 July 2026)

# La collezione di strumenti didattici di fisica dell'Istituto Maria SS.ma del Rosario di Palermo

**Aurelio Agliolo Gallitto**

*Dipartimento di Fisica e Chimica - Emilio Segrè, University of Palermo, Palermo, Italy*

Email: aurelio.agliologallitto@unipa.it

**RIASSUNTO**

Nell'autunno del 2025, l'Istituto Maria SS.ma del Rosario (IMSSR) di Palermo ha donato al Dipartimento di Fisica e Chimica - Emilio Segrè dell'Università di Palermo una collezione di strumenti didattici di fisica. Gli strumenti sono stati identificati, classificati in sezioni ed ora esposti all'interno della Collezione Storica degli Strumenti di Fisica, nell'edificio storico di via Archirafi 36. La raccolta, composta da circa sessanta strumenti prevalentemente destinati all'insegnamento della fisica classica nelle scuole superiori dell'Istituto, costituisce una significativa testimonianza della cultura sperimentale della fisica scolastica nella prima metà del Novecento. L'articolo ne ricostruisce l'origine storica, ne illustra gli ambiti disciplinari e le principali tipologie strumentali, e ne analizza il valore scientifico e didattico in una prospettiva di museologia scientifica descrittiva e storica, con particolare attenzione al ruolo dei cataloghi commerciali quali strumenti di mediazione didattica e tecnica.

**Nota:** L'articolo è presentato in versione bilingue inglese-italiano. Il testo inglese è seguito dalla versione originale italiana.

# Introduzione

L’Istituto Maria SS.ma del Rosario di Palermo (IMSSR), gestito dalle Suore Domenicane del S. Cuore di Gesù, fu fondato il 19 giugno 1925 presso il monastero domenicano di S. Caterina a Palermo. Inizialmente, fu istituito un corso completo di scuola elementare e media inferiore. A causa del conflitto mondiale, le attività furono interrotte e la sede venne temporaneamente trasferita presso il palazzo dei Principi di Cutò in via Lincoln a Palermo. Nel 1948 venne acquistato il terreno in via Gian Filippo Ingrassia 44 dove fu edificata la nuova sede. Con il trasferimento nel 1949 nella nuova sede, le iscrizioni aumentarono e si aprirono vari indirizzi, con doppie sezioni: Liceo Classico, Istituto Tecnico Femminile, Perito Aziendale Corrispondenze in Lingue estere (PACLE). Vennero quindi ampliati i laboratori didattici e si acquistarono numerosi strumenti scientifici. Nello stesso periodo, il Collegio ospitava più di 150 ragazze provenienti dai paesi limitrofi. Nel corso degli anni, tuttavia, la gestione dell’Istituto divenne progressivamente più complessa, portando alla chiusura degli indirizzi della scuola secondaria superiore. Nell’autunno del 2025, a seguito della cessazione delle attività laboratoriali, l’Istituto ha donato gli strumenti didattici di fisica e chimica al Dipartimento di Fisica e Chimica - Emilio Segrè (DiFC) dell’Università di Palermo.

Gli strumenti sono stati identificati, classificati in sezioni ed esposti in due vetrine coeve all’interno della Collezione Storica degli Strumenti di Fisica (sito web 1), situata nell’edificio storico di via Archirafi 36. La figura 1 mostra le due vetrine espositive. La raccolta, composta da circa sessanta strumenti prevalentemente destinati all’insegnamento della fisica classica, è stata denominata Collezione IMSSR. L’elenco completo degli strumenti, corredato da una breve descrizione, è consultabile nella pagina web (sito web 2).

**Figura 1.** Vetrine espositive collocate nell’edificio storico di via Archirafi 36 del DiFC, Università di Palermo.

La Collezione IMSSR costituisce una significativa testimonianza della cultura sperimentale della fisica scolastica nella prima metà del Novecento documentando una fase storica in cui l’insegnamento della fisica nelle scuole secondarie italiane si fondava in larga misura sull’uso sistematico di apparati sperimentali standardizzati, selezionati sulla base dei cataloghi commerciali (Frick, 1856; Pike, 1856; Fumeo, 1930 ca; Paravia, 1938 & 1953; Magini, 1940; Vallardi, 1960). Tali apparati riflettono una didattica fortemente sperimentale, in linea con la tradizione della fisica classica delineata nei manuali ottocenteschi e mantenuta nei programmi scolastici fino alla metà del Novecento (Ganot, 1863; Jamin, 1870; Murani, 1917).

Il presente articolo illustra gli ambiti disciplinari e le principali tipologie strumentali della Collezione IMSSR. Esso analizza alcuni strumenti di particolare interesse come fonti documentarie rilevanti per la ricostruzione della storia dell’insegnamento della fisica, un ambito meno indagato rispetto alla strumentazione destinata alla ricerca scientifica. In prospettiva museologica, tali collezioni consentono di esaminare il rapporto tra sapere scientifico, didattica e produzione industriale, mettendo in evidenza il ruolo dei cataloghi commerciali come strumenti di mediazione didattica e tecnica.

## Ambiti disciplinari della Collezione

Gli strumenti sono stati classificati secondo criteri disciplinari coerenti con la tradizione didattica della fisica classica. Il settore maggiormente rappresentato è quello della meccanica, con strumenti destinati allo studio del moto, dell'equilibrio e delle macchine semplici: piani inclinati, carrucole e bilance dinamometriche. La figura 2 mostra l'apparato per il parallelogramma delle forze, con due carrucole e quattro aste graduate. L'apparato è descritto da Frick nel catalogo del 1856 (Frick, 1856: 58) ed è illustrato nella figura 3.

**Figura 2.** Apparato per il parallelogramma delle forze, con due carrucole e quattro aste graduate, produzione Vallardi, ca. 1950.

**Figura 3.** Schema dell'apparato per il parallelogramma delle forze, con due carrucole e quattro aste graduate, tratto da (Frick, 1856: 58).

La sezione dedicata allo studio dei fluidi comprende strumenti destinati alla dimostrazione delle principali leggi dell'idrostatica e della pneumatica, ampiamente utilizzati nell'insegnamento della fisica nelle scuole secondarie tra la fine dell'Ottocento e la prima metà del Novecento. Tra le tipologie maggiormente rappresentate figurano strumenti per lo studio della pressione nei liquidi, della spinta idrostatica e del principio di Archimede, nonché apparecchi per la dimostrazione della trasmissione della pressione nei fluidi e dei fenomeni legati alla pressione atmosferica. Nella figura 4 è mostrata una bilancia idrostatica con doppio cilindro di Archimede; nella stessa figura è visibile un areometro di Nicholson a volume costante, utilizzato per la determinazione del peso specifico di corpi solidi insolubili in acqua. L'uso di questo strumento è descritto nell'estratto del catalogo Pike (Pike, 1856: 227) riportato nella figura 5. Tali dispositivi consentivano di visualizzare concetti astratti attraverso configurazioni semplici e robuste, progettate specificamente per un uso ripetuto in ambito scolastico.

**Figura 4.** Bilancia idrostatica elementare con doppio cilindro di Archimede montata su una base di legno, con giogo elevabile e regolabile mediante vite di pressione: comprende tre piatti di ugual peso, di cui uno munito di gancetto per l'applicazione del doppio cilindro di Archimede (Catalogo G. B. Paravia 1953, N. 50120, Lire 9800). Sul lato destro si vede anche l'Areometro di Nicholson a volume costante, per la determinazione del peso specifico di corpi solidi insolubili in acqua.

**Figura 5.** Descrizione e illustrazione dell'areometro di Nicholson tratta da (Pike, 1856).

La sezione di termologia comprende strumenti destinati allo studio dei fenomeni termici, della dilatazione e della trasmissione del calore. La figura 6 mostra una cassetta di Ingenhousz, per il confronto della conducibilità termica di 5 sostanze diverse (alluminio, ottone, plexiglas, acciaio, legno). Si menziona la presenza in questa sezione di un raro termometro a minima e massima tipo Six-Bellani (Jamin, 1870: 105), inventato nel 1782 da James Six (1731-1793) e perfezionato all'inizio del XIX secolo da Angelo Bellani (1776-1852).

**Figura 6.** Cassetta di Ingenhousz, per il confronto della conducibilità termica di 5 sostanze diverse (alluminio, ottone, plexiglas, acciaio, legno), rappresentate da sbarrette applicabili a una cassetta di lamierino, produzione Fumeo, ca. 1950.

Gli strumenti di acustica – tra cui il metronomo di Maelzel-Paquet, un diapason su cassetta di risonanza e un apparecchio per lo studio della propagazione del suono nel vuoto – testimoniano un approccio didattico incentrato sull'esperienza sensoriale e sulla relazione tra fenomeno fisico e percezione uditiva. Queste tipologie strumentali trovano ampio riscontro nei manuali e nei cataloghi per l'insegnamento della fisica tra Otto e Novecento (Blaserna, 1875).

La sezione di ottica comprende strumenti impiegati per lo studio dei fenomeni fondamentali della propagazione della luce, della riflessione e della formazione delle immagini. Gli apparati conservati riflettono un'impostazione didattica fortemente orientata all'ottica geometrica. Infatti, particolare rilievo assumono gli strumenti destinati alla dimostrazione delle leggi della riflessione, sia in configurazioni semplici sia in apparati più complessi, progettati per visualizzare fenomeni quali la riflessione multipla e la dipendenza dell'immagine dall'orientamento delle superfici riflettenti. L'uso di specchi piani, sferici concavi e convessi consentiva di affrontare in modo intuitivo concetti quali il fuoco e la formazione dell'immagine reale e virtuale. Strumenti di carattere più divulgativo e dimostrativo, come il caleidoscopio, contribuivano inoltre a stimolare l'interesse degli studenti attraverso l'osservazione di effetti ottici e geometrici immediatamente osservabili.

La sezione dedicata al magnetismo e all'elettrologia comprende strumenti impiegati per lo studio dei fenomeni magnetici ed elettrici attraverso dimostrazioni prevalentemente qualitative. La sezione di elettrologia documenta l'insegnamento dei fenomeni di elettrizzazione, accumulo e scarica elettrica, nonché delle prime nozioni di circuiti elettrici. Strumenti quali elettroscopi, elettrofori e bottiglie di Leida erano largamente impiegati per dimostrare la presenza di cariche elettriche e i meccanismi di induzione, offrendo esperienze spettacolari ma didatticamente efficaci. Accanto a questi, dispositivi per la realizzazione di semplici circuiti e generatori elettrici testimoniano l'introduzione progressiva dell'elettricità applicata nella didattica scolastica. Per quanto riguarda il magnetismo, sono presenti strumenti destinati alla rivelazione e allo studio dei campi magnetici, dell'orientamento degli aghi magnetici e delle interazioni tra magneti e correnti elettriche. Bussole, aghi magnetici e dispositivi dimostrativi consentivano di illustrare in modo intuitivo concetti quali il campo magnetico, la polarità e l'azione magnetica della corrente. Apparati più sensibili, come i galvanometri con il sistema astatico, permettevano inoltre di rendere apprezzabili correnti di debole intensità, introducendo gli studenti allo studio quantitativo dei fenomeni elettromagnetici. A questo scopo, il sistema astatico dell'apparecchio mostrato nella figura 7 descrive il cosiddetto "moltiplicatore" del galvanometro astatico di Nobili ideato nel 1825 dal fisico Leopoldo Nobili (1784 - 1835), da cui lo strumento prende il nome (sito web 3). Nobili per eliminare l'effetto del campo magnetico terrestre usò due aghi identici, rigidamente montati paralleli sullo stesso asse, ma con i poli magnetici orientati in versi opposti, appesi allo stesso filo di seta. Il sistema del doppio ago rende quindi il galvanometro più sensibile di quelli a singolo ago magnetico. Nella figura 8 è illustrato lo schema astatico tratto da (Ganot, 1863):

> "*… consiste in due aghi sovraposti parallelamente, prospicienti i poli contrari (fig. 653), l'uno interno, l'altro esterno al circuito, e rannodati in modo da non poter girare l'uno senza l'altro […] l'effetto principale del sistema astatico è di ridurre l'azione direttrice della Terra*".

**Figura 7.** Sistema astatico del galvanometro di Nobili, produzione Fumeo, ca. 1950.

**Figura 8.** Schema del sistema astatico con moltiplicatore del galvanometro di Nobili (Ganot, 1863: 602).

Infine, la sezione di chimica ed elettrochimica comprende strumenti destinati all'insegnamento sperimentale dei fondamenti della chimica generale e dei fenomeni elettrochimici, utilizzati nei laboratori scolastici della prima metà del Novecento. Gli apparati conservati testimoniano un approccio didattico fortemente basato sulla manipolazione diretta, sull'osservazione dei fenomeni e sulla riproduzione controllata di reazioni chimiche ed elettrolitiche. La figura 9 mostra l'apparecchio di Kipp, ampiamente utilizzato in laboratorio per la produzione di idrogeno, acido solfidrico e anidride carbonica mediante opportune reazioni chimiche tra una soluzione acquosa acida e una sostanza solida insolubile (un metallo o un sale). L'apparecchio fu realizzato per la prima volta nel 1844 dal chimico olandese Petrus Jacobus Kipp (1808-1864); esso permette in generale di produrre vari gas, in quantità controllabile, da un'opportuna reazione chimica tra una soluzione acquosa acida e un solido insolubile.

Particolare rilievo assumono gli strumenti concepiti per l'apprendimento della nomenclatura chimica e della composizione delle sostanze, che riflettono l'esigenza di rendere accessibili concetti astratti attraverso supporti materiali e modelli combinatori. In questo contesto si colloca la cassetta di nomenclatura chimica “FORMULATOR”, contenente 72 dischi di legno, ciascuno dei quali reca su una faccia il simbolo dell'elemento chimico rappresentato, con una codifica cromatica differenziata in base alla valenza. I dischi sono collegabili tra loro mediante apposite asticciole metalliche elastiche, che consentono la costruzione e la visualizzazione di formule chimiche anche complesse (Catalogo Paravia 1953, n. 50265, lire 6800). Tali dispositivi evidenziano l'importanza attribuita, nella didattica scolastica dell'epoca, alla visualizzazione delle relazioni tra elementi chimici e alla costruzione delle formule come esercizio logico e manuale.

La sezione include inoltre apparati destinati allo studio dei processi elettrochimici, quali la produzione di corrente elettrica mediante reazioni chimiche e la decomposizione elettrolitica delle soluzioni. Strumenti come pile e voltametri consentivano di illustrare il legame tra chimica ed elettricità, offrendo esempi concreti di trasformazione dell'energia e introducendo gli studenti ai principi fondamentali dell'elettrochimica.

Accanto a tali strumenti, la presenza di vetreria da laboratorio testimonia l'attenzione riservata all'insegnamento delle tecniche sperimentali di base. In particolare, dispositivi per la produzione di gas, come l'apparecchio di Kipp (Erdmann, 1900; sito web 4), mostrato nella figura 9, consentivano di svolgere dimostrazioni fondamentali anche sulle proprietà dei gas, aspetti centrali nella formazione scientifica scolastica.

**Figura 9.** Apparecchio di Kipp. Realizzato nel 1844 dal chimico olandese Petrus Jacobus Kipp (1808 - 1864) per la produzione di gas (quali idrogeno, anidride carbonica, acido solfidrico) per reazione tra un acido e un metallo o un sale.

Un aspetto di particolare rilievo della Collezione IMSSR è la stretta corrispondenza tra gli strumenti conservati e i cataloghi commerciali delle ditte produttrici di materiale didattico scientifico. Cataloghi quali quelli di Amedeo Fumeo, Giovan Battista Paravia e Antonio Vallardi non costituivano meri strumenti di vendita, ma svolgevano un ruolo centrale come mediatori culturali, contribuendo a orientare le scelte didattiche e a standardizzare l'equipaggiamento dei laboratori scolastici. Nella figura 10 sono mostrate le targhette delle case produttrici presenti sugli strumenti. Attraverso la selezione, la descrizione e la classificazione degli apparati, tali cataloghi concorrevano alla definizione delle pratiche educative, suggerendo modalità d'uso, sequenze sperimentali e ambiti disciplinari di applicazione. Essi rappresentano pertanto una preziosa fonte documentaria per lo studio del rapporto tra produzione industriale, programmi scolastici e didattica sperimentale, nonché per l'analisi dei processi di circolazione e consolidamento del sapere scientifico in ambito scolastico. In questa prospettiva, la Collezione IMSSR consente di mettere in relazione diretta gli strumenti scientifici con le fonti editoriali che ne hanno promosso la diffusione, offrendo una lettura integrata della cultura materiale della didattica scientifica nella scuola italiana della prima metà del Novecento.

**Figura 10.** Targhette delle case produttrici presenti sugli strumenti della Collezione IMSSR: (a) Amedeo Fumeo, Società per Azioni, Milano; (b) Giovan Battista Paravia & C., Materiale didattico scientifico, Torino; (c) Antonio Vallardi, Materiale scientifico, Milano; (d) Antonio Vallardi, Editore, Milano.

## Discussione e conclusioni

Gli strumenti della Collezione IMSSR riflettono un'impostazione didattica fondata sull'osservazione diretta dei fenomeni fisici e sulla dimostrazione sperimentale delle leggi fondamentali, secondo una tradizione consolidata nella pratica scolastica tra la fine dell'Ottocento e la prima metà del Novecento. L'analisi delle diverse sezioni disciplinari evidenzia modelli didattici comuni, basati sull'uso di apparati semplici, robusti e standardizzati, concepiti per un impiego ripetuto nei laboratori scolastici.

Il caso della Collezione IMSSR contribuisce a colmare una lacuna nella letteratura relativa alla strumentazione scientifica scolastica, spesso trascurata rispetto a quella accademica. Dal punto di vista storico-didattico, gli strumenti della Collezione documentano una fase in cui l'insegnamento della fisica nelle scuole secondarie italiane si fondava in larga misura sull'adozione sistematica di strumenti di laboratorio, prodotti in serie e distribuiti attraverso i cataloghi commerciali delle principali ditte specializzate nel materiale didattico scientifico. La stretta corrispondenza tra gli esemplari conservati e le descrizioni catalografiche testimonia il ruolo centrale svolto da tali cataloghi nella definizione delle pratiche educative e nell'omogeneizzazione dell'equipaggiamento dei laboratori scolastici. In questa prospettiva, la Collezione IMSSR si configura come un'importante fonte documentaria per la ricostruzione della cultura sperimentale della fisica scolastica, mettendo in luce il legame tra sapere scientifico, didattica sperimentale e produzione industriale. Il passaggio degli strumenti dal contesto d'uso originario a quello museale consente oggi una lettura storica e comparativa, valorizzandoli non solo come supporti didattici, ma come documenti della storia dell'educazione scientifica.

Il trasferimento della Collezione dall'ambito scolastico a quello museale ne ha pertanto modificato lo statuto, trasformando gli strumenti da oggetti funzionali a beni culturali di interesse scientifico. La loro integrazione all'interno della Collezione Storica degli Strumenti di Fisica dell'Università di Palermo (Sear, 2017; Agliolo Gallitto et al, 2018) consente di collocarli in un quadro storico-scientifico più ampio, in relazione con la strumentazione universitaria coeva e con le pratiche didattiche e di ricerca sviluppate in ambito accademico (Barbacci et al, 2012; Agliolo Gallitto et al, 2017). In tale contesto, la Collezione IMSSR costituisce un nucleo coerente e organico all'interno della Collezione Storica degli Strumenti di Fisica dell'Ateneo, capace di restituire una visione articolata dell'insegnamento della fisica e delle scienze affini nella scuola italiana della prima metà del XX secolo, evidenziando al contempo continuità e differenze tra didattica scolastica e didattica universitaria. La donazione ha contribuito in modo significativo all'arricchimento della collezione universitaria, che oggi comprende circa seicento strumenti scientifici e didattici a partire dall'inizio del XIX secolo (Sear, 2017; Agliolo Gallitto et al., 2018; sito web 1).

Dal punto di vista museologico, la valorizzazione della Collezione IMSSR risponde pienamente agli obiettivi della museologia scientifica, in particolare per quanto riguarda la tutela, l'interpretazione e la trasmissione della cultura materiale della scienza. L'inserimento nel patrimonio universitario ha reso possibile non solo la conservazione degli strumenti, ma anche la loro contestualizzazione storica, la catalogazione sistematica e la fruizione a fini di ricerca, didattica e divulgazione.

In conclusione, la Collezione IMSSR rappresenta un significativo esempio di collezione storica di strumenti scientifici di origine scolastica. Il suo studio consente di integrare la storia della fisica, la storia dell'educazione scientifica e la museologia descrittiva e storica, offrendo nuove prospettive per la ricerca sul patrimonio scientifico e per la riflessione sulle pratiche di conservazione e valorizzazione delle collezioni didattiche in ambito universitario.

## Ringraziamenti

L'autore desidera ringraziare Suor Rosa Maria La Bella per aver reso possibile la donazione; Maria Rosalia Carotenuto, Salvatore Cataldo e Fabio Panfili per i proficui suggerimenti; Marcello Mirabello per il supporto tecnico; Germana Mulè e Francesca Tignola per il contributo fornito nella ricerca di fonti bibliografiche storiche.

## Bibliografia

## Siti web (ultimo accesso 22 luglio 2026)